\documentclass[letterpaper,10pt,twocolumn]{article}
\usepackage{ol3}
\usepackage{graphicx,amsmath,amssymb,epstopdf,cite,subfig}
\usepackage{leitian}

\begin{document}

\twocolumn[ 

\title{Compressive X--ray phase tomography based on the transport of intensity equation}

\author{Lei Tian,$^{1,*}$ Jonathan C. Petruccelli,$^{1}$ Qin Miao,$^{1}$ Haris Kudrolli,$^2$ Vivek Nagarkar,$^2$and \mbox{George Barbastathis$^{1,3}$}}
\address{$^1$Department of Mechanical Engineering, Massachusetts Institute of Technology, Cambridge, MA 02139, USA \\
$^2$RMD, Inc., Watertown, MA, USA \\
$^3$Singapore-MIT Alliance for Research and Technology (SMART) Centre, Singapore 138602, Singapore\\
$^*$Corresponding author: {lei\_tian@alum.mit.edu}
}

\begin{abstract}
We develop and implement a compressive reconstruction method for tomographic recovery of refractive index distribution for weakly attenuating objects in a microfocus X--ray system.  This is achieved through the development of a discretized operator modeling both the transport of intensity equation and X--ray transform that is suitable for iterative reconstruction techniques.
\end{abstract}
 ] 

Traditional tomography with hard X--rays recovers the attenuation of an object. Attenuation does not always provide good contrast when imaging objects made of materials with low electron density, {\it e.g.} soft tissues. In these cases, richer information is often contained in the phase, {\it i.e.} the optical thickness of the sample~\cite{momose1996phase, pfeiffer2006phase, davis1995phase}. Propagation based techniques are particularly suitable for X--ray phase imaging because they allow phase to be recovered from intensity images taken at multiple propagation distances without the need for optical elements~\cite{Guigay:07, Chen:13}. Here we adopt the transport of intensity equation (TIE) which relates the measured intensity to the Laplacian of the phase under a weakly--attenuating sample approximation.  Implementing TIE at many angles while rotating the object allows tomographic reconstruction of the refractive index distribution. 

TIE tomographic reconstruction first requires a suitable forward model, consisting of \mbox{1) a} projection of refractive index through the sample and \mbox{2) modeling} diffraction after the sample with the TIE.  Recovering the phase then amounts to inverting these operations on the measured data.  A straightforward method of reconstruction is to invert the forward model in two steps~\cite{Burvall:11}.  For the first step, the TIE can be solved by a Poisson equation solver.  The TIE is ill--posed because the transfer function relating the intensity measurement to the phase tends to zero as the spatial frequency decreases.  As a result, reconstructions are often corrupted by significant low--frequency noise, requiring regularization.  Tikhonov regularization is most commonly employed to reduce these artifacts~\cite{Gureyev:98, Burvall:11}.  For the second step, a standard tomographic reconstruction is carried out, e.g. using the filtered back--projection (FBP) method.  In order for FBP to yield a result free from high--frequency ``streaking'' artifacts, projections must be taken at many angles.  It is often desirable to use fewer projections in order to reduce dose or acquisition time, in which case the tomographic inversion problem is underdetermined. Rather than solve these two inverse problems independently, the forward and inverse models may be adapted into a single--step operation combining TIE and tomography~\cite{Bronnikov:02, Groso:06, 2006SPIE.6318E..30G, Baillie:12}.  However, a single step inversion still requires many projection angles in order to avoid artifacts.  Iterative solvers have recently been proposed to reduce these artifacts when attempting a reconstruction from a small number of projections.  Myers {\it et. al.} propose inversion of TIE tomography measurements to obtain a sample distribution using prior knowledge that the sample consists of  a single material of known refractive index~\cite{Myers:08}.  Sidky {\it et. al.} propose inverting only the tomographic measurement to obtain a boundary--enhanced image~\cite{Sidky:10}.  A similar approach has also been proposed to combine a different phase retrieval technique [contrast transfer function (CTF)] and algebraic tomographic reconstruction~\cite{Kostenko:13oeTV}. In this Letter, we design a forward model that combines TIE and tomography operations in a single, discretized linear operator in the Fourier domain and develop an iterative reconstruction method for recovery of refractive index of a weakly--attenuating object.

\begin{figure}[t]
\centering
\includegraphics[width=0.45\textwidth]{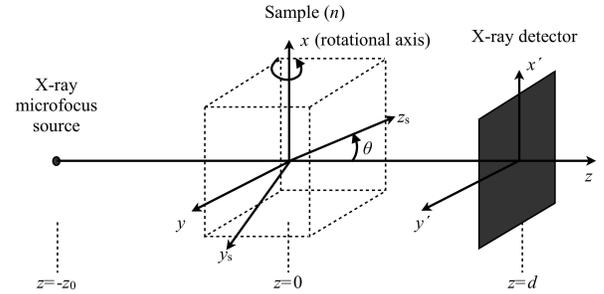}
\caption{Imaging geometry for TIE tomography}
\label{setup}
\end{figure}

A schematic diagram of the imaging geometry is shown in Fig.~\ref{setup}. A point source with mean wavelength $\lambda$ is located at the plane of $z=-z_0$. This is a good approximation for a table--top microfocus X--ray source. The (weakly attenuating) sample is characterized by a real--valued refractive index \mbox{$n(x,y,z)$}, where the origin of the cartesian coordinates is located within the object.  We further assume that the the object is small enough that the beam passing through it can be approximated as a plane wave oriented along the $z$ axis and that the interaction between the sample and the field can be treated using the projection approximation, {\it i.e.} phase delay imparted upon the field passing through the sample is \mbox{$\phi=(2\pi/\lambda) \int  n(x,y,z)\md z$}.  The intensity $I$ is recorded by an area detector located at $(x',y',d)$.  Although the TIE for microfocus sources is usually formulated using two measurements of intensity at different positions along the optical axis~\cite{Gureyev:98}, because the sample is assumed to be weakly attenutating,  we consider a fixed detector position with two images taken with and without the sample in place, $I$ and $I_{\mi}$, respectively. Under the paraxial and small--wavelength approximations, the relationship between $I$ and $\phi$ is given by
\be
\label{xrayTIE}
g(x,y;\theta) \!\equiv\! \frac{2\pi}{\lambda d'}\left[1\!-\!\frac{I(M_{\rs}x, M_{\rs}y)}{I_{\mi}(M_{\rs}x, M_{\rs}y)}\right]\!=\! \delxy^2\phi(x,y;\theta),\!\!
\ee
where $\delxy$ is the gradient operator in the $(x,y)$ plane, $M_{\rs}=(z_0+d)/z_0$ is a magnification factor, and $d' = z_0 d/(z_0+d)$ is the effective propagation distance. Equation~\ref{xrayTIE} is a finite difference form of the TIE that uses two images taken with and without the sample in place in order to recover the projected phase of a weakly attenuating sample.  Taken together, Eqs.~\eqref{xrayTIE} and~\eqref{pure} specify the continuous model describing how the refractive index of a weakly attenuating sample creates visible signature in the intensity under the assumption that the object subtends a small solid angle as viewed from the source.  In a tomographic measurement, the sample is rotated about the $x$--axis such that the projected phase $\phi$ for a given rotation angle $\theta$ is
 \be
\phi(x,y;\theta)\!=\!\!\iint \!n(x,y_{\rs},z_{\rs})\delta(y\!-\!y_{\rs}\!\cos\theta\!+\!z_{\rs}\!\sin\theta)\md y_{\rs}\md z_{\rs},\!\!.
\label{pure}
\ee

In order to construct an iterative process for the retrieval of phase from measured intensity, one must first represent these two operations in a suitable discretized form.  Let us assume our detector consists of an $N\times N$ grid of square pixels of side length $M_\mathrm{s}\Delta$ and let $\Theta$ denote the number of angular projections.  Then the measured $g$ at all projection angles may be arranged into a real--valued vector $\bg$ of length $N^2\Theta$.  The refractive index of the object may be discretized into a 3D volume consisting of $N^3$ cubic voxels of side length $\Delta$ and arranged into a vector $\bn$.  Let $\bP$ denote an $N^2\Theta\times N^2\Theta$ matrix representing the discretized form of Eq.~\eqref{xrayTIE} and $\bR$ denote the $N^2\Theta\times N^3$ matrix representing Eq.~\eqref{pure} such that
\be
\bg = \bP\bR\:\bn \equiv \mathbf{A} \bn,
\label{fwdmodel}
\ee 
where $\mathbf{A}$ represents a matrix combining both $\bP$ and $\bR$ into a single operation.  

An effective approach to invert this equation and solve for $\bn$ if our data is undersampled is to adapt compressed sensing theory by assuming $\bn$ can be expressed as sparse, {\it i.e.} it contains only a small number of nonzero coefficients in some specified basis~\cite{candes2006robust}. Since the sample of interest often consists of regions with constant refractive indices, we choose the following compressive reconstruction model
\be
\hat{\bn}=\mathop{\mathrm{arg~min}}_{\bn}\|\bn\|_{\TV} \mbox{ such that } \bg = \mathbf{A}\bn,
\label{CTIETomo}
\ee
where the TV function $\|\bn\|_{\TV}$ is our sparsity basis, defined as
\mbox{$\|\bn\|_{\TV} = \sum\sqrt{(\nabla_x \bn)^2
+({\nabla_y \bn})^2
+({\nabla_z \bn})^2}$}, and $\nabla_x$, $\nabla_y$, and $\nabla_z$ are the finite difference operators in the three Cartesian spatial dimensions. We adapt the two--step iterative shrinkage/thresholding algorithm (TwIST)~\cite{2007ITIP...16.2992B} to solve the minimization.

\begin{figure*}[tbh]
\centering
{\includegraphics[width=0.8\textwidth]{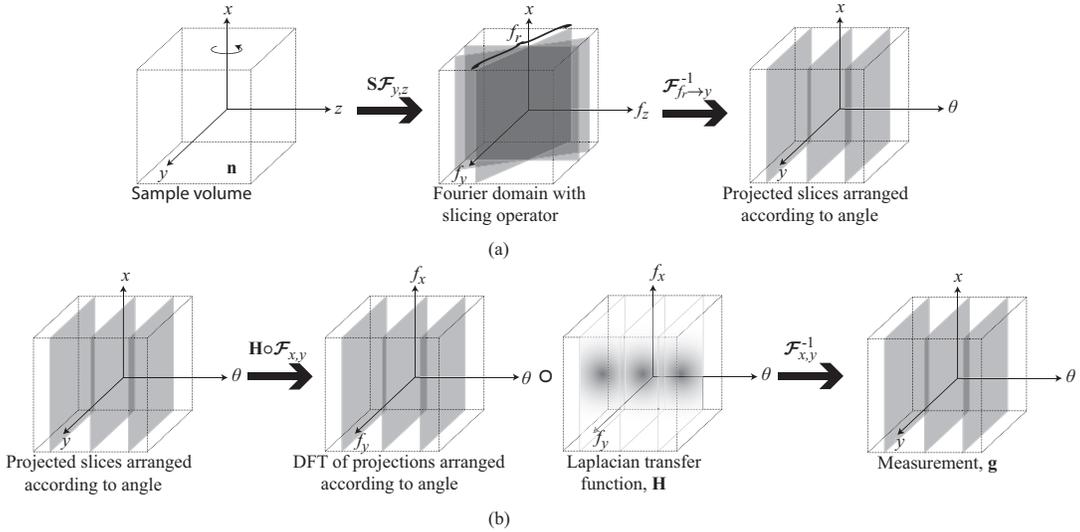}}
\caption{Illustration of the construction of (a) the discrete projection operator $\bR$ and (b) the discrete TIE operator $\bP$.}
\label{Operators}
\end{figure*}

Although in its final form $\bA$ takes the vector $\bn$ to the measurement vector $\bg$, it is more natural to think about construction of $\bA$ in terms of $\bn$ and $\bg$ arranged as matrices of dimensions $N\times N\times N$ and $N\times N\times\Theta$, respectively, as illustrated in Fig.~\ref{Operators}.  For notational simplicity, rather than referring to the entries of these matrices by their indices, we use the discretized values of position or spatial frequency associated with that entry.  In what follows we denote the 2D discrete Fourier transform (DFT) matrix which applies a DFT from the Cartesian coordinates $e_1$ and $e_2$ to spatial frequencies $f_{e_1}$ and $f_{e_2}$ as $\FT_{e_1,e_2}$ and its inverse DFT (IDFT) as $\FT^{-1}_{e_1,e_2}$. 

The discrete projection operator $\bR$ can be implemented in either the spatial domain~\cite{Myers:08, Sidky:10, Kostenko:13oeTV} or the Fourier domain~\cite{fessler2003nonuniform}. We adopt the Fourier domain method here since it is robust to discretization error and noise, and computationally much more efficient~\cite{matej2004iterative}.  The Fourier transform of each projection can be computed using the projection--slice theorem: 1) $\FT_{y,z}$ is applied to $\bn$, 2) 
radial slices through the $x$--axis, each corresponding to a given projection angle, are taken in the Fourier domain by an operator $\mathbf{S}$, 3) the spatial distributions of projected phase over all projection angles are obtained by a 1D DFT over these slices from the (radial) spatial frequency coordinate $f_r$ to the spatial coordinate $y$, an operation denoted by $\FT_{f_r\rightarrow y}^{-1}$.  The steps in this construction are illustrated graphically in Fig.~\ref{Operators}(a). 

The TIE operator $\bP$ is simply the discrete Laplacian operator acting on the projected phases, which can be implemented efficiently in the Fourier domain as 
\be
\bP = \delxy^2 = \FT_{x,y}^{-1} \mathbf{H}\circ\FT_{x,y},
\ee
where the transfer function $\mathbf{H}$ is an \mbox{$N\times N\times\Theta$} matrix with each \mbox{$N\times N$} slice along the $\theta$ dimension identical to \mbox{$-4\pi^2(f_x^2+f_y^2)$}, and $\circ$ denotes entry--wise multiplication.

There is a subtlety regarding the implementation of these discrete operators to ensure that the output $\bg$ has the correct dimensionality and sample spacing, and that the operator produces results that are stable under iteration.  First, recall that $\bg$ is constructed from measurements over two spatial dimensions at a set of projection angles such that the spatial coordinates are sampled over a Cartesian grid with sample spacing $\Delta$.  Recall also that we define $\bn$ over voxels of side length $\Delta$.  All DFTs and IDFTs are implemented over Cartesian grids in space or spatial frequencies of sample spacing $\Delta$ and $(N\Delta)^{-1}$, respectively, which allows implementation via fast Fourier transforms (FFT).  However, attention to resampling is required when implementing the operator $\mathbf{S}$.  In order to construct a grid over $(x,f_r,\theta)$ from planar slices through data on the grid $(x,f_y,f_z)$, $\mathbf{S}$ must incorporate interpolation and then resampling (gridding).  We implement this by adapting the non-uniform FFT (NUFFT) algorithm~\cite{fessler2003nonuniform} such that the NUFFT operator $\FT_{\text{NU}}=\mathbf{S}\FT_{y,z}$.  The TIE and projection operators may then be combined and simplified (eliminating a 1D DFT--IDFT pair) as 
\be
\mathbf{A} = \FT_{x,y}^{-1} \mathbf{H}\circ\FT_{x}\FT_{\text{NU}}.
\label{eq:ForwardOp}
\ee
$\FT_{\text{NU}}$ produces an output with the proper grid spacing and ensures the stability required for the iterative algorithms utilized in compressive reconstruction.

\begin{figure}[b]
\centering
\includegraphics[width=0.35\textwidth]{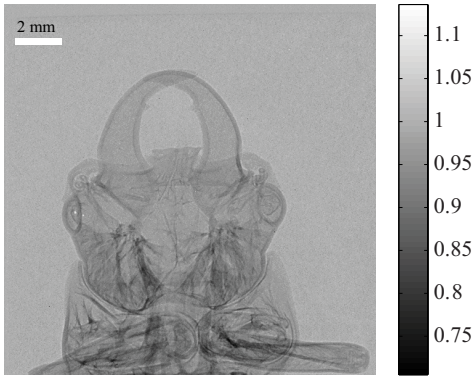}
\caption{Single projected measurement of $I$.}
\label{I_beetle}
\end{figure}

\begin{figure*}[tbh]
\centering
{\includegraphics[width=\textwidth]{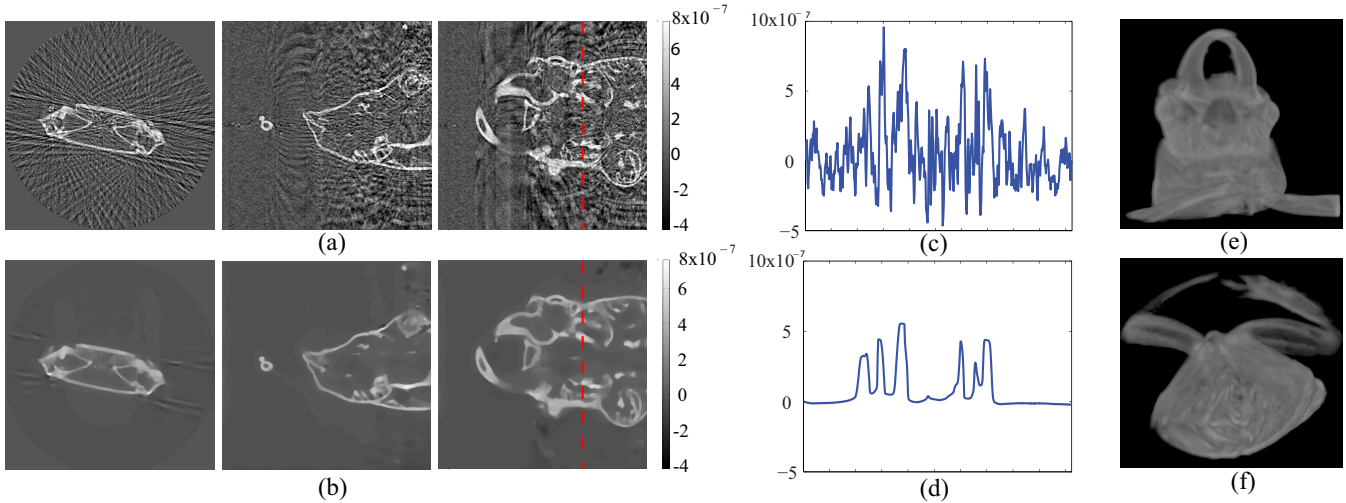}}
\caption{Reconstruction results for the refractive index deviation from air $(1-n)$. (a) Fourier based TIE solver + FBP; (b) Compressive reconstruction; (c) Plot along the dashed line illustrated in (a); (d) Plot along the dashed line illustrated in (b).  The three cross--sections are taken from the three orthogonal planes going through the center of the sample; (e--f) Two perspectives of a 3D rendering of the reconstruction in (b).   [Online: movies show the rotation of the renderings in (e) (Media 1) and (f) (Media 2) about their respective vertical axes.]}
\label{phi_beetle}
\end{figure*}

A microfocus source (Hamamatsu L8121--03) located at $z_0 = 0.765$m, was operated at 40kVp and 100mA to produce a circular focal spot of 5$\mu$m in diameter.  The resulting X--ray beam has central wavelength $\lambda = 0.062$nm. Intensity images were taken with a custom designed EMCCD based X--ray camera, where a 150$\mu$m thick RMD CsI:Tl scintillator was fiber optically coupled to an EMCCD [Andor, ixon, 512$\times$512 pixels ($N=512$), 16$\mu$m pixel size] with a 6:1 taper~\cite{nagarkar2006fast}; the effective pixel size at the scintillator is 96$\mu$m. The scintillator was at $d=1.711$m ($M_{\rs}=3.24$, $\Delta\approx 30\mu$m).  For a beetle sample, a single image was taken at every 5 degrees ($\Theta=36$) with 6.7 seconds exposure time during the tomographic measurement,.  The discretized sample $\bn$ therefore consists of $512^3$ voxels, while the data vector $\bg$ has $512^2\times36$ entries.  The intensity of the incident beam $I_{\mi}$ was calibrated by taking a single background image without the sample in place.  A single projection measurement of $I$ is shown in Fig.~\ref{I_beetle}, which shows the range of attenuation through the sample.  Notice that the maximum attenuation of the incident light is approximately $70\%$ of its initial value, which is not strictly negligible.  However, it has been previously observed that phase retrieval using the weak--attenuation formula is quite robust to values of attenuation even beyond those seen in this sample~\cite{Groso:06,Myers:08}.

During the reconstruction, we first compute $g$ for each angle. Reconstruction results using two different methods are compared in Fig.~\ref{phi_beetle} for the refractive index deviation from the surrounding air $(1-n)$.  In (a), the Fourier domain TIE solver with Tikhonov regularization chosen to provide optimal results is used to compute phase projections at each angle, and then the FBP method with a Ram--Lak filter is applied for the tomographic inversion. The results suffer from severe streaking artifacts due to missing samples between slices in the Fourier domain. Low--frequency artifacts (blurring) around edges are also observable but are less severe as compared to a single projected phase reconstruction. This is likely due to denser sampling around the origin resulting from the intersection of the Fourier slices. Both artifacts can be greatly suppressed using compressive reconstruction with TV minimization, Eq.~\eqref{CTIETomo}, whose results are shown in (b), since TV minimization favors large structures with sharp edges. Plots of the $1-n$ along the dashed lines in (a) and (b) are illustrated in Fig.~\ref{phi_beetle}(c) for the FBP method and in (d) for the compressive reconstruction.  Notive the significant reduction in high frequency artifacts in from (c) to (d).  3D renderings of the compressive reconstruction of the refractive index are shown in Fig.~\ref{phi_beetle}(c) and (d).

Compressed sensing works best when the measurement is ``incoherent,'' {\it i.e.} the sparse information in the unknown is evenly spread out in the measurement~\cite{candes2006robust}.  This is achieved in our model by the projection operator $\bR$.  The incoherence of the measurement could be improved through the use of source coding~\cite{JP:13_COSI} or coded aperture~\cite{MacCabe:12} which is beyond the scope of our current work.

This work was supported by the Department of Homeland Security, Science and Technology Directorate through contract HSHQDC-11-C-00083. We are grateful to David J. Brady, Justin Lee and Rajiv Gupta for helpful discussions.

\end{document}